\documentclass[fleqn,usenatbib]{mnras}

\usepackage{newtxtext,newtxmath}

\usepackage[T1]{fontenc}

\DeclareRobustCommand{\VAN}[3]{#2}
\let\VANthebibliography\thebibliography
\def\thebibliography{\DeclareRobustCommand{\VAN}[3]{##3}\VANthebibliography}

\usepackage{graphicx}	
\usepackage{amsmath}	
\usepackage{multicol}
\usepackage{ulem}

\usepackage{color}
\usepackage[dvipsnames]{xcolor}

\usepackage{tikz,hyperref}

\definecolor{lime}{HTML}{A6CE39}
\DeclareRobustCommand{\orcidicon}{
	\begin{tikzpicture}
	\draw[lime, fill=lime] (0,0) 
	circle [radius=0.16] 
	node[white] {{\fontfamily{qag}\selectfont \tiny ID}};
	\draw[white, fill=white] (-0.0625,0.095) 
	circle [radius=0.007];
	\end{tikzpicture}
	\hspace{-2mm}
}

\foreach \x in {A, ..., Z}{\expandafter\xdef\csname orcid\x\endcsname{\noexpand\href{https://orcid.org/\csname orcidauthor\x\endcsname}
			{\noexpand\orcidicon}}
}

\def \kms{\ifmmode{~{\rm km\,s}^{-1}}\else{~km~s$^{-1}$}\fi}

\title[The barium central star of the PN Abell 70]{A detailed study of the barium central star of the planetary nebula Abell 70}
\author[D. Jones et.\ al]{David Jones$^{1,2}\orcidA{}$ \thanks{E-mail:
djones@iac.es}, Henri~M.~J. Boffin$^{3}\orcidB{}$, Alex J. Brown$^{4}\orcidC{}$,
Jiri Zak$^{3}\orcidD{}$, George Hume$^{4,5}\orcidE{}$,
\newauthor James Munday$^{6}\orcidF{}$,
 and Brent Miszalski$^{7}\orcidG{}$
\\
$^{1}$Instituto de Astrof\'isica de Canarias, E-38205 La Laguna, Tenerife, Spain\\
$^{2}$Departamento de Astrof\'isica, Universidad de La Laguna, E-38206 La Laguna, Tenerife, Spain\\
$^{3}$European Southern Observatory, Karl-Schwarzschild Strasse 2, 85748 Garching, Germany\\
$^{4}$Department of Physics and Astronomy, University of Sheffield, Sheffield, S3 7RH, UK\\
$^{5}$Isaac Newton Group of Telescopes, Apartado de Correos 368, E-38700 Santa Cruz de La Palma, Spain\\
$^{6}$Department of Physics, Gibbet Hill Road, University of Warwick, Coventry CV4 7AL, United Kingdom\\
$^{7}$Australian Astronomical Optics - Macquarie, Faculty of Science and Engineering, Macquarie University, North Ryde, NSW 2113, Australia\\
}

\date{Accepted xxxx xxxxxxxx xx. Received xxxx xxxxxxxx xx; in original form xxxx xxxxxxxx xx}

\pagerange{\pageref{firstpage}--\pageref{lastpage}} \pubyear{2022}

\begin{document}
\label{firstpage}
\pagerange{\pageref{firstpage}--\pageref{lastpage}}
\maketitle

\begin{abstract}
We present a detailed study of the barium star at the heart of the planetary nebula Abell 70.  Time-series photometry obtained over a period of more than ten years demonstrates that the barium-contaminated companion is a rapid rotator with temporal variability due to spots.  The amplitude and phasing of the photometric variability changes abruptly, however there is no evidence for a change in the rotation period (P = 2.06~d) over the course of the observations.  The co-addition of 17 high-resolution spectra obtained with VLT-UVES allow us to measure the physical and chemical properties of the companion, confirming it to be a chromospherically-active, late G-type sub-giant with more than +1~dex of barium enhancement. We find no evidence of radial velocity variability in the spectra, obtained over the course of approximately 130~d with a single additional point some 8 years later, with the radial velocities of all epochs approximately $-$10 \kms{} from the previously measured systemic velocity of the nebula. This is perhaps indicative that the binary has a relatively long period (P $\gtrsim$ 2~yr) and high eccentricity ($e\gtrsim$ 0.3), and that all the observations were taken around radial velocity minimum. However, unless the binary orbital plane is not aligned with the waist of the nebula or the systemic velocity of the binary is not equal to the literature value for the nebula, this would imply an unfeasibly large mass for the nebular progenitor.
\end{abstract}


\section{Introduction}
\label{sec:intro}

It is now clear that binary stars hold the key to understanding the formation of many planetary nebulae \citep[PNe;][]{jones17a}. However, with much of the recent focus being placed on common envelope evolution \citep{boffin19}, the role of wider binaries remains almost completely unconstrained \citep{tyndall13}.  To date, only three wide binary central stars of PNe have confirmed periods \citep{jones17b} with other systems either being so wide that they are resolved \citep{ciardullo99} or displaying composite spectra with giant companions \citep{tyndall13}.  Four central star systems are known to be barium stars \citep{bond03,miszalski12a,miszalski13a,2019A&A...624A...1L} with only one having a known orbital period \citep[LoTr~5;][]{jones17b,aller18}.  Barium stars are mostly -- but not only: see, e.g., \cite{2019A&A...626A.128E} -- giant stars of spectral type G-K which display enhanced abundances of carbon and s-process elements like barium and strontium, and are now known to all be long-period binaries \citep[100d $\lesssim$ P$_\mathrm{orb}$ $\lesssim$ 10$^4$d;][]{mcclure80,2019A&A...626A.127J}.  The chemical contamination of barium stars is believed to be due to accretion of chemically enriched material from an evolved binary companion, likely through wind \citep[][]{boffin88} or wind Roche lobe overflow \citep[WRLOF;][]{theuns93,nagae04}.  WRLOF will occur when the acceleration radius of the asymptotic giant branch (now a white dwarf in these barium star systems) star's stellar wind is comparable or greater than its Roche lobe radius \citep{mohamed12}, meaning that the wind itself is strongly influenced by the binary potential and can be accreted onto the companion via the first Lagrangian point.  This accreted material chemically contaminates the surface of the companion as well as increases its spin rate due to the conservation of angular momentum \citep{theuns96}.  Intriguingly, the amount of material required to account for the observed contamination is often in excess of that which would be expected to result in critical rotation rates in the companion, providing an indication that significant (and as yet not understood) angular momentum losses must be experienced in these systems \citep{matrozis17}.  PNe with barium central stars offer an important window into this process as the presence of the short-lived nebula \citep[$\tau\leq$ 30\,000 yr;][]{jacob13} means that the mass transfer has occurred too recently for significant changes in the companion's spin rate, and moreover the nebula itself traces the mass-loss history of the system as it is formed from the material that has escaped the binary potential \citep{jones18a}.

In this paper, we present a detailed study of the central star of the planetary nebula Abell 70 (PN~G038.1$-$25.4, or A66~70, hereafter A~70), which was revealed by \cite{miszalski12a} to be a binary comprised of a  hot white dwarf, revealed by GALEX photometry, and a roughly G8IV-V  barium star secondary, which has been heavily polluted by s-process material from the nebular progenitor ([Ba/Fe] overabundance $\sim$0.5 dex).  A~70 itself is a southern PN most well known for its striking appearance, earning it the nickname the ``diamond ring" due to its ring-like form and superposition with a background elliptical galaxy \citep[which forms the diamond of the diamond ring,][]{miszalski12a}.
The kinematical study of \cite{tyndall13} showed A~70 to most likely comprise a faint bipolar shell, the waist of which is encircled by a bright toroid, the deprojection of which results in an estimated nebular inclination of 30\degr $\pm$ 10\degr{}.  They also determined a systemic heliocentric velocity for the nebula of $-$73$\pm$4 \kms{} and a kinematical age of approximately 15,000 years for the distance of 6.04$\pm$2.13 kpc derived by \cite{frew16}.

The central star of Abell 70 appears in the Gaia Data Release 3 \citep[DR3; source id: 6907822573352460032,][]{gaiadr3}, where it was not identified as binary nor was it bright enough for the radial velocity data to be released.  However, Gaia DR3 does contain stellar parameters for the central star of A~70 \citep{gaia-apsis1,gaia-apsis2}, derived using the General Stellar Parameterizer from Photometry (GSP-Phot) module of the Astrophysical parameters inference system (Apsis). GSP-Phot simultaneously fits the BP/RP spectra, parallax and apparent magnitude, using isochrone models to constrain the stellar properties.  The corresponding parameters, as well as the geometric parallax from EDR3 \citep{bailer-jones21}, for the central star of A~70 are summarised in Tab.\ \ref{tab:gaia}, however these should be used with caution as will be discussed in Sec.\ \ref{sec:param}.

\section{Photometry}
\label{sec:a70phot}

\subsection{New i-band photometry}
\label{sec:a70photnew}

We supplement the literature photometry of \citet{bond18} with photometric observations of the central star of A~70 made in the $i$-band at various facilities between August 2010 and July 2022.

The first of the supplementary observations were acquired on the nights of 2010 August 11--13 and 15--17 with the 1.9-m Radcliffe Telescope of the South African Astronomical Observatory (SAAO) and the SAAO CCD instrument.  The STE4 CCD was employed along with the Bessell I filter, providing a FOV of approximately 2.5\arcmin{}$\times$2.5\arcmin{} and a binned pixel scale of 0.28\arcsec{} pixel$^{-1}$.  

Further observations were then acquired on the nights of 2013 June 3--7 with the European Southern Observatory's 3.6-m New Technology Telescope (ESO-NTT) and the ESO Faint Object Spectrograph and Camera v.2 (EFOSC2) instrument \citep{EFOSC2a,EFOSC2b}.  EFOSC2 was employed in its imaging mode along with a Gunn $i$-band filter (ESO ID \#705) to provide a FOV of 4.1\arcmin{}$\times$4.1\arcmin{} with a binned pixel scale of 0.24\arcsec{} pixel$^{-1}$.  

Service mode observations were acquired with the Las Cumbres Observatory Global Telescope \citep[LCOGT;][]{brown13} on 2015 April 1, May 14, and June 12, 14, 18, 19, 20, 21, 22, 25 and 27 with the 1m network, and on 2015 June 24 and 27  on the 2m network.  All observations were taken with a Cousins I-band filter, with 1m network observations employing SBIG cameras for a 15.8\arcmin{}$\times$15.8\arcmin{} FOV and an unbinned pixel scale of 0.23\arcsec{} pixel$^{-1}$, while the 2m network observations employed Spectral cameras for a 10.5\arcmin{}$\times$10.5\arcmin{} FOV and a binned pixel scale of 0.30\arcsec{} pixel$^{-1}$.

Finally, observations were taken with the 2-m Liverpool Telescope (LT) and the IO:O instrument on the nights of 2022 May 25, 27, 28, 29 and 31, June 1, 11, 12, 17 and 18, and July 4, 7 and 16.  All observations were taken unbinned (10\arcmin{}$\times$10\arcmin{} FOV and 0.15\arcsec{} pixel$^{-1}$) and with a Sloan $i'$ filter.

All data were bias and flat field corrected using standard reduction routines or the instrument pipelines (where available).  Differential aperture photometry of the central star of A~70 was then performed using the astropy-affiliated \texttt{photutils} python package\footnote{\url{https://photutils.readthedocs.io/en/stable/}}. 

\begin{figure*}
\centering
\includegraphics[width=\textwidth]{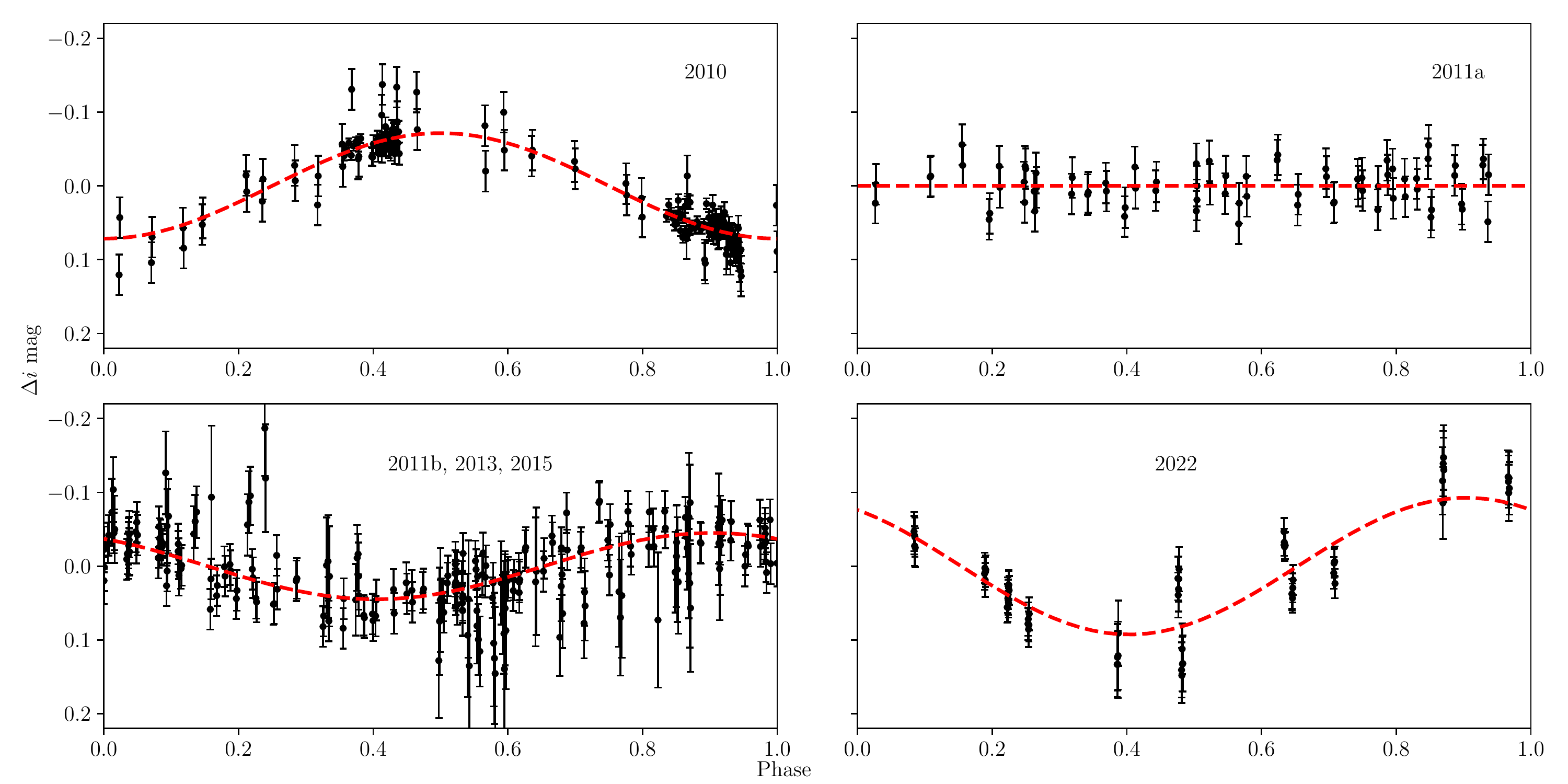}
\caption{The combined $i$-band photometry \citep[including the data from][]{bond18} demonstrating the constant periodicity (P=2.061 d), but changing phasing and amplitude, of the variability.  All data are phased according to the ephemeris derived from the 2010 data using the period derived from the 2011b, 2013 and 2015 data.}
\label{fig:a70phot}
\end{figure*}

\begin{figure}
\centering
\includegraphics[width=\columnwidth]{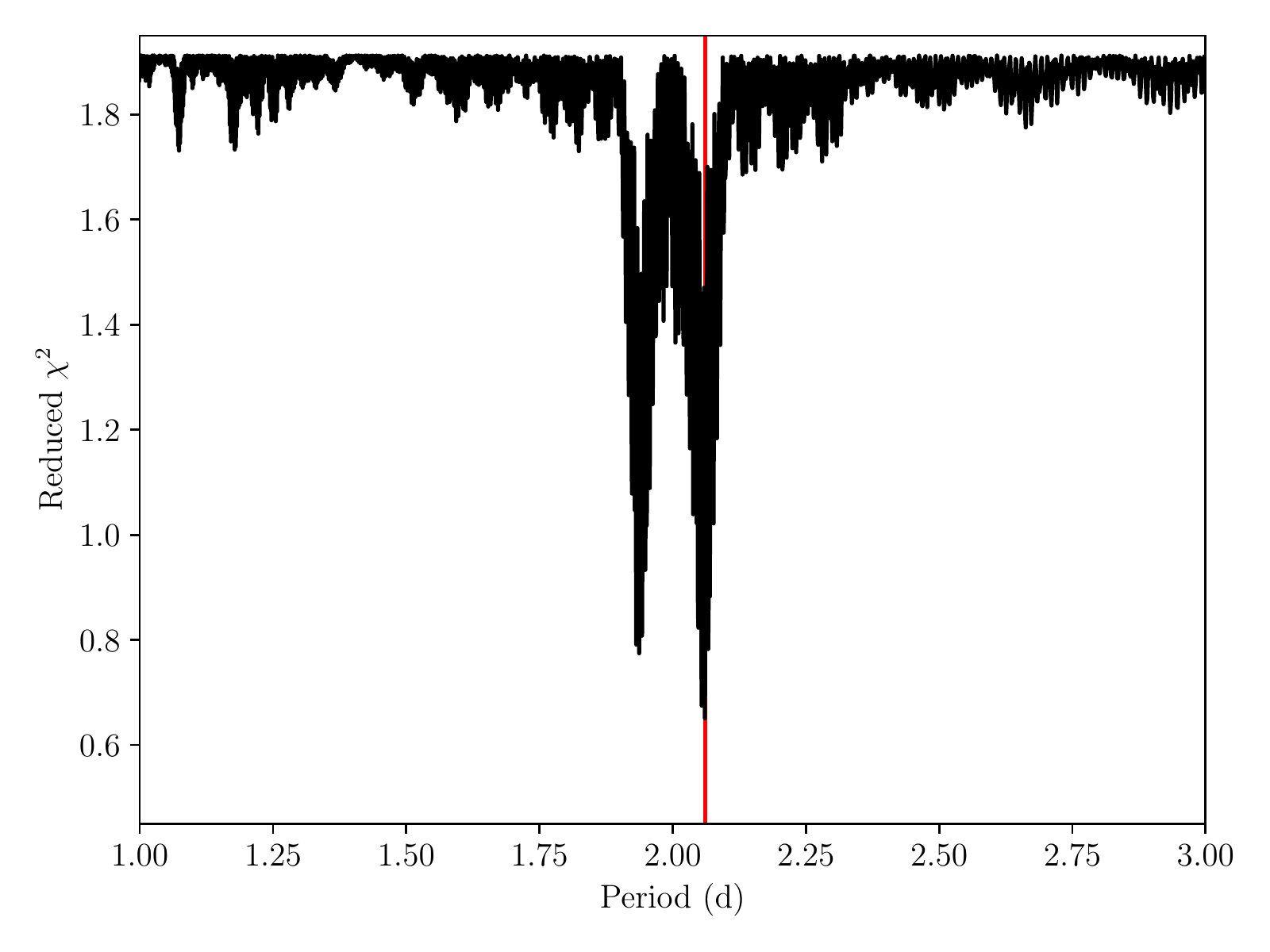}
\caption{The reduced $\chi^2$ of a sinusoidal fit as a function of period for the 2011b, 2013 and 2015 data (shown folded on this period in Fig.\ \ref{fig:a70phot}.  The best fitting period (P=2.061$\pm$0.005d) is marked by the vertical red line.}
\label{fig:a70period}
\end{figure}

\citet{bond18} identify a periodicity of 2.06~d in their data, with a similar period clearly evident in the newly acquired data.  However, no single period and time of first minima could be found that fits all the data (perhaps not unexpected given that the variability is clearly not constant, see Sec.\ \ref{sec:varphot} for further discussion).  In order to better constrain the period, as well as the changes between observing epochs, we focus on the periodicitiy of all the data taken between 2011 and 2015 (shown in the lower left panel of Fig.\ \ref{fig:a70phot}).  Given the nature of the variability, we measure the period via the reduced $\chi^2$ of a sinusoidal fit as a function of frequency \citep[ideally suited to sinusoidal variability, e.g.,][]{horne86b,munday20}.  The resulting periodogram is shown in Fig.\ \ref{fig:a70period}.  Two strong minima in the reduced $\chi^2$ are found, the strongest at 2.061$\pm$0.005~d \citep[consistent with the period of ][]{bond18} and a second slightly shallower at 1.938$\pm$0.005~d.  The data from 2010, combining that of \citet{bond18} and our own from SAAO, phases well on the first period but not the second, lower period, likely indicating that this is an alias.  As such, we conclude that the rotation period is indeed $\approx$2.06~d and that there is no strong evidence for a change in period before and after the phase of negligible variability in early 2011.  There is, however, a clear shift in phase between the 2010 data and the later data used to derive the period (see Fig.\ \ref{sec:a70phot}).  There is no statistically significant evidence for a shift between the 2011--2015 data and the later data from 2022, although the lack of data in the intervening seven years as well as the uncertainty on the period mean that we cannot exclude a phase shift between these two data sets.  The 2022 data does, however, appear to have a larger photometric amplitude, more consistent with the 2010 data.  Nonetheless, if the variability did remain constant throughout this time, we can refine the period to P=2.060561$\pm$0.000002~d.

\subsection{Survey photometry}

The field containing A~70 has been observed over relatively long periods of time by a number of all-sky surveys.  Unfortunately, the relatively faint central star and bright nebula combined with the large pixel scales employed by the majority of these surveys means their photometry is not particularly useful in directly probing the variability in A~70 -- the 2~d period would not be uniquely identified in any of the datasets.  Nonetheless, they can be used to probe the apparent changes in phase between the epochs presented in Fig.\ \ref{fig:a70phot}. The details of this analysis are outlined in Appendix \ref{sec:surveyphot}, with the main conclusion being that we cannot rule out an additional shift in phase between the 2015 and 2022 epochs of photometry.

\subsection{Origin of the changing variability}
\label{sec:varphot}

The observed photometric period -- presumably the rotation period of the less evolved component of the binary nucleus of A~70 -- is consistent with the periodicities observed in other fast-rotating companions to central stars of PNe.  \cite{miszalski13a} found a period of 5.5 d for Hen~2-39, while similar periods have been found for giant components of  WeBo~1 \citep[4.7~d;][]{bond03}, LoTr~1 \citep[6.4~d;][]{tyndall13} and LoTr~5 \citep[5.95~d;][]{aller18}.  The study of \cite{miszalski12a} found that the companion in A~70 is either a main sequence or sub-giant G-type star, both of which would be below break-up for a rotational period of 2~d (although a sub-giant would be fairly close).  

However, no other rapidly rotating companion has been shown to present with changing variability.  Within observing periods, where the data is sufficiently precise \citep[e.g., the 2022 LT data or the 2010 data from ][]{bond18}, there seems to be some evidence for additional scatter beyond the over-arching sinusoidal periodicity. This may simply be underestimated uncertainties which do not account for the contamination from the surrounding nebula, however the sinusoidal variability itself also seems to come and go as well as change in phasing. In early 2011, the variability drops below a detectable level, and then returns in late 2011 with a different phasing but similar amplitude.  The new data obtained in 2013 and 2015 phase well with the late 2011 data, indicating that the variability was relatively stable over this four year period. The long time span between the 2015 and 2022 observations means that we cannot conclusively determine whether there has been a phase shift during that time (small uncertainties in the rotation period equate to significant changes in phase).  Nonetheless, similar shifts in phase of photometry variability are observed in other magnetically active stars (not at the centres of PNe), for example the giant component of IT Com where the change was associated with a periastron passage \citep{olah13}. The fact that the central star of A~70 is the first of the rotationally variable PN central stars to be found to show changes in its photometric variability is more likely a consequence of the lack of long-term monitoring for other central stars rather than a special property of the central star of A~70.

\begin{table}
\caption{The Gaia DR3 parameters of the central star of A~70.}
\label{tab:gaia}
\centering
\begin{tabular}{ll}
\hline
Parameter & A~70\\
\hline
$T_\mathrm{eff}$ (K)		&	5144$^{+19}_{-17}$\\
log $g$					& 4.784$^{+0.012}_{-0.015}$\\
{[Fe/H]}					&$-2.73^{+0.26}_{-0.32}$\\
GSP-Phot Aeneas distance (kpc)        & 1.47$^{+0.02}_{-0.01}$\\
GSP-Phot Aeneas radius (R$_\odot$) & 0.506$^{+0.05}_{-0.04}$\\
Parallax, $\varpi$  (mas) & 0.25$\pm$0.12\\
Geometric parallax distance (kpc) & 3.6$^{+1.5}_{-1.0}$\\
Bp magnitude & 18.13$\pm$0.03\\
G magnitude & 17.66$\pm$0.01\\
Rp magnitude & 17.08$\pm$0.02\\
\hline
\end{tabular}
\end{table}

\section{Spectroscopy}
\label{sec:a70spec}
The central star of A~70 was observed 16 times between 2011 May 28 and September 4, and once more on 2019 October 17 (for exact dates see Tab.\ \ref{tab:a70rvs}) using the Ultraviolet and Visual \'Echelle Spectrograph \citep[UVES;][]{dekker00} on ESO's Very Large Telescope (VLT) Unit Telescope 2 (UT2), also known as Kueyen.  All exposures were of 3000~s through a 1\arcsec{} slit. The dichroic \#2 was employed, with the red arm centred at 7600\AA{} (with a BK7\_5 blocking filter) and the blue arm centred at 4370\AA{} (with a HER\_5 below-slit filter).  This set-up provides continuous spectral coverage in the ranges 3800-5000\AA{} (blue-arm) and 5700-9400\AA{} (red-arm, with a small gap in coverage between CCDs around 7600\AA{}), with a resolution of approximately 40\,000. The data were reduced using the standard ESO pipeline.

\citet{miszalski12a} presented intermediate-dispersion spectroscopy obtained with the Gemini Multi-Object Spectrograph \citep[GMOS;][]{hook04} on the Gemini South Telescope, and with the FOcal Reducer and low-dispersion Spectrograph \citep[FORS2;][]{FORS} on ESO's VLT Unit Telescope 1 (UT1, Antu).  In order to analyse them with the same methodology as the aforementioned UVES spectra, these data (as well as a further two FORS2 spectra both obtained on 2012 July 8 with the same instrumental setup) were downloaded from their respective archives and re-reduced.

\subsection{Radial velocities}
\begin{figure}
\centering
\includegraphics[width=\columnwidth]{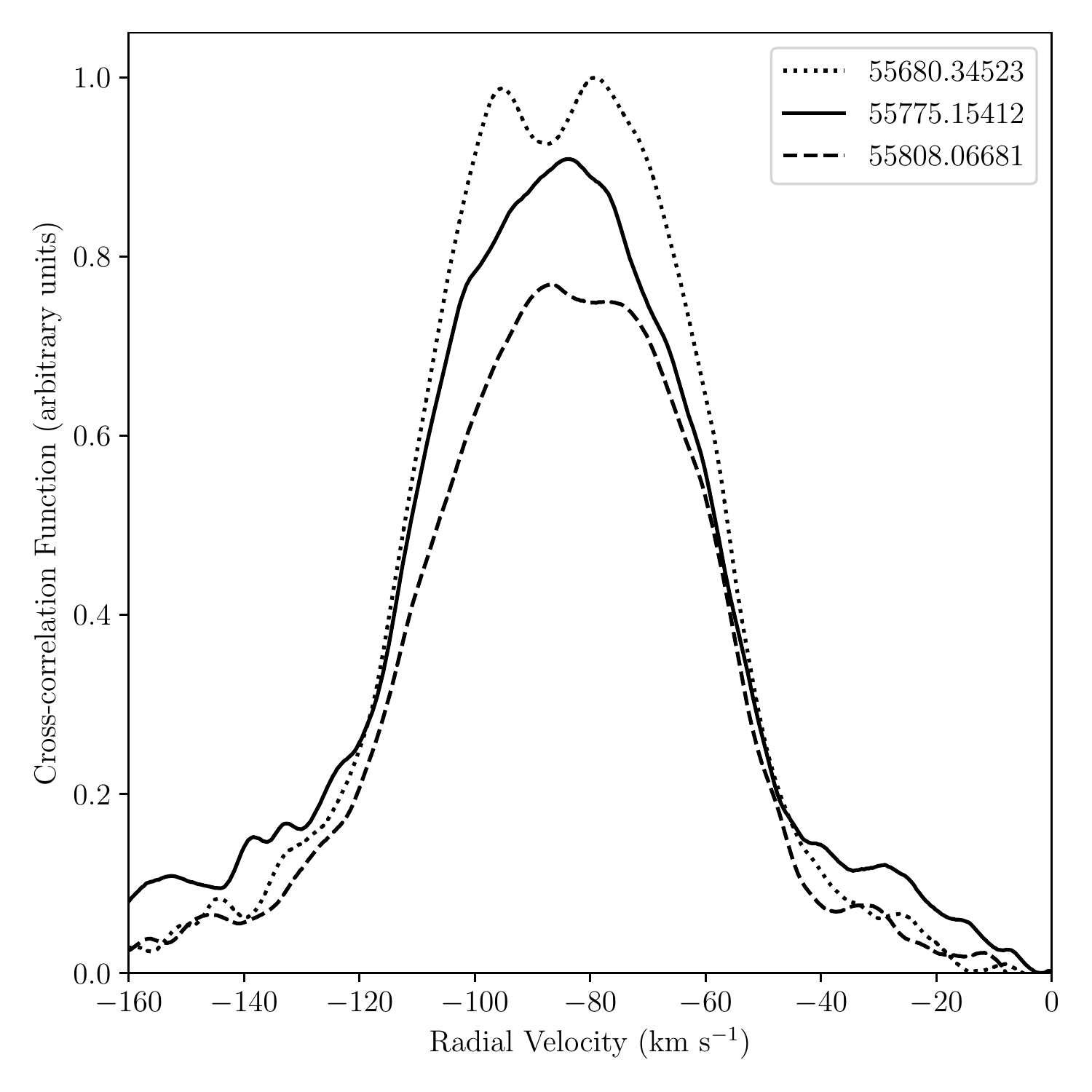}
\caption{The CCFs of three spectra obtained close to photoemtric minimum (the MJDs of the spectra are in the legend, but each is $\approx$0.1 in phase away from the photometric minimum), highlighting that the shape of the CCF does not appear to correlate with the photometric phase.}
\label{fig:CCF}
\end{figure}

The reduced spectra were continuum subtracted, corrected to heliocentric velocity and 
cross-correlated against a G8IV template \citep[in accordance with the spectral type determined by][]{miszalski12a} created using \texttt{spectrum}\footnote{ \url{http://www.appstate.edu/~grayro/spectrum/spectrum.html} } \citep{spectrum}.  Each cross-correlation function (CCF) displayed a clear peak of width consistent with a high $v$ sin $i$ ($\sim$30--50 \kms{}) as expected from the rotational period derived in section \ref{sec:a70phot}.  For the highest signal-to-noise spectra the shape of the CCF could be seen to deviate from a single gaussian peak, perhaps indicative of spots crossing the surface of the star (as expected from the observed photometric variability).  This hypothesis is difficult to test given the relatively sparse phase sampling, as well as the low signal-to-noise of the individual spectra and, particularly, individual lines - thus preventing a detailed doppler tomography of the stellar surface.  In any case, the shape of the CCF does not seem to correlate with the photometric phase of the observation (see Fig.\ \ref{fig:CCF}).

In order to minimise the effects of the possible spot(s) on the shape of the CCF, the final radial velocities were measured using a bisector of the entire peak, although some ``jitter'' may still be present.  The resulting heliocentric radial velocities (RVs) are presented in Tab.\ \ref{tab:a70rvs}.  Note that just as in \citet{miszalski12a}, the FORS2 spectra showed significant signs of flexure, and so the derived RVs were corrected using gaussian fits to the [O~\textsc{iii}] 4959\AA{} and 5007\AA{} emission lines and assuming a nebular systemic velocity of $-$73 \kms{} \citep{tyndall13}.

\begin{figure}
\centering
\includegraphics[width=\columnwidth]{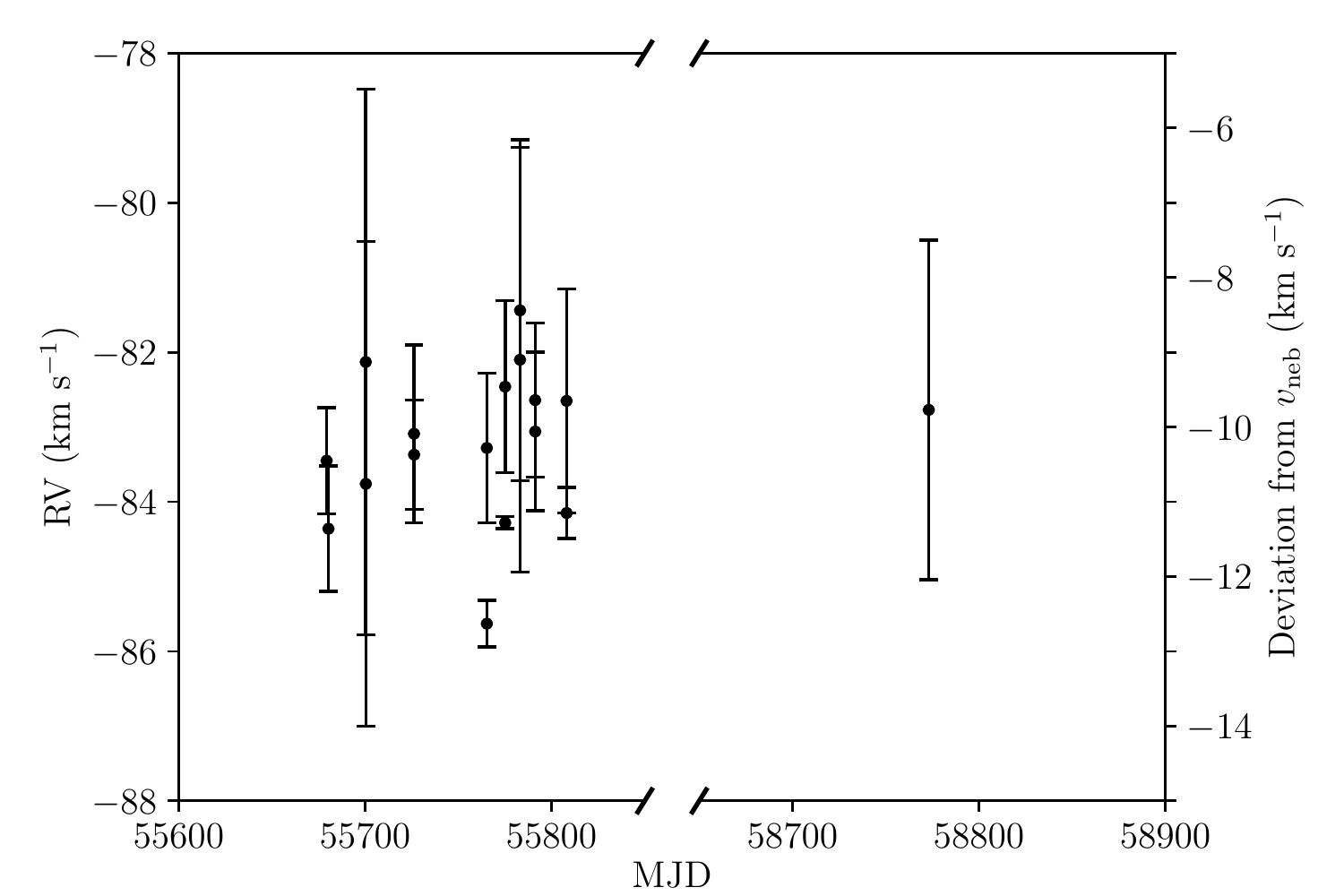}
\caption{Heliocentric radial velocity measurements of the central star of A~70.  The right-hand axis shows the deviation of the stellar heliocentric radial velocity from the heliocentric systemic radial velocity of the nebula as measured by \citet{tyndall13}.}
\label{fig:a70rvs}
\end{figure}

\begin{figure}
\centering
\includegraphics[width=\columnwidth]{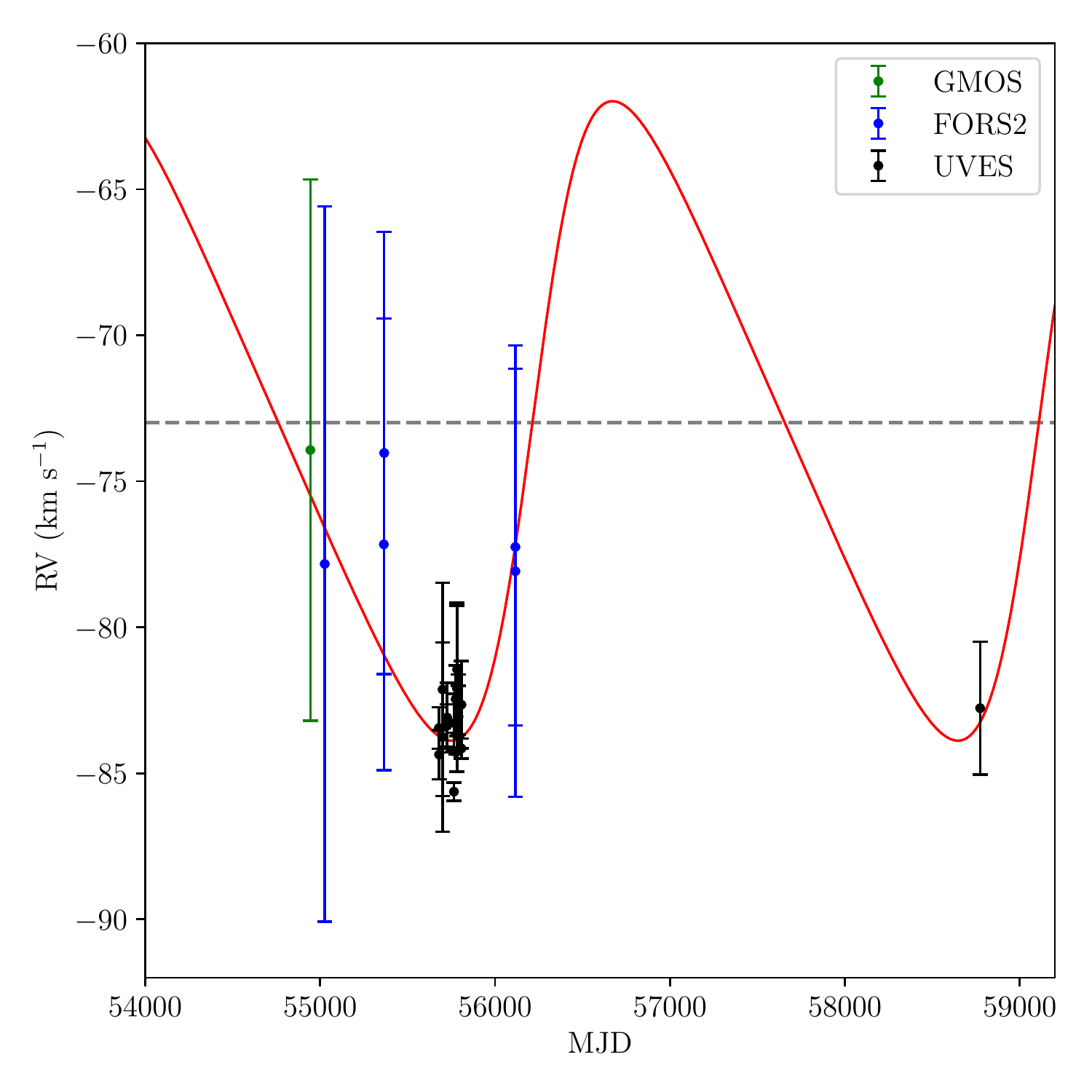}
\caption{A speculative orbital solution with a period of approximately 8 years and an eccentricity of 0.3. The horizontal dashed line is the nebular velocity derived by \citet{tyndall13}, taken to be the systemic velocity of the binary for each of the fits.
}
\label{fig:orbit}
\end{figure}

The resulting RVs show no obvious periodic variability, with the UVES measurements clustering around a heliocentric velocity of between $-$85 and $-$82 \kms{} (See Fig.\ \ref{fig:a70rvs}.  Given that the typical periods of barium stars are in the range of hundreds to tens of thousands of days \citep{jorissen98,jorissen19}, one would perhaps expect to see some signs of variability in the UVES RVs. The 2011 data spans the shorter end of the barium star period distribution (P$\sim$100 days), such that periods of a year or shorter can be ruled out (unless the orbital inclination is particularly low).  However, with only a single data point some 8 years after the initial 2011 data, a number of orbital solutions are still possible.  If one reasonably assumes that the nebular systemic velocity, $-$73 \kms{} \citep{tyndall13}, is also the systemic velocity of the binary, then the semi-amplitude of the binary would be roughly 10 \kms{} with all data being taken close to quadrature (i.e., at the RV minimum).  The lack of observed variability would then be a consequence of a relatively long period and high eccentricity.  Restricting to just the UVES data, periods of approximately 2, 4 and 8 years would all give plausible fits to the data.  However, including the generally more uncertain GMOS and FORS2 radial velocities clearly favours a longer period (see Fig.\ \ref{fig:orbit}).  This longer period would also be supported by the previously discussed photometric variability where the variability was found to be stable for at least four years (see Sec. \ref{sec:a70phot}).  Note, however, that the similar changes in photometric variability of other magnetically-active stars may have been associated with periastron passages, perhaps implying that the time of pericenter passage for the central stars of A~70 should be roughly coincident with the 2011 period of negligible photometric variability.  This is not the case for the fit presented in Fig.\ \ref{fig:orbit} and no reasonable fit could be obtained constraining the time of pericenter passage.  Ultimately, given the low precision of the GMOS and FORS2 RVs, and the non-variability of the UVES RVs, any fit is highly speculative.

\begin{table}
\caption{A table of measured heliocentric radial velocities versus heliocentric Julian date for the polluted giant secondary of A~70.}
\label{tab:a70rvs}
\centering
\begin{tabular}{lrllc}
\hline
 Heliocentric  & \multicolumn{2}{c}{Radial velocity} & Instrument & Approx.\ S/N\\
  MJD (days)  & \multicolumn{2}{c}{(\kms{})}&& of spectrum\\
\hline
54944.32710 & $-$73.93 & $\pm$9.27 & GMOS    & 9.3 \\
55026.41908 & $-$77.83 & $\pm$12.25 & FORS2  & 4.0  \\
55364.42193 & $-$77.16 & $\pm$7.73 & FORS2   & 10.5 \\
55366.41773 & $-$74.03 & $\pm$7.57 & FORS2   & 12.8 \\
55679.35887 & $-$83.45 & $\pm$0.71 & UVES    & 5.4 \\
55680.34523 & $-$84.36 & $\pm$0.84 & UVES    & 7.1 \\
55700.36627 & $-$82.13 & $\pm$3.65 & UVES    & 2.8 \\
55700.40375 & $-$83.76 & $\pm$3.24 & UVES    & 2.7 \\
55726.23343 & $-$83.09 & $\pm$1.19 & UVES    & 4.3 \\
55726.29200 & $-$83.37 & $\pm$0.73 & UVES    & 4.0 \\
55765.28905 & $-$83.28 & $\pm$1.00 & UVES    & 2.6 \\
55765.32868 & $-$85.63 & $\pm$0.31 & UVES    & 2.3 \\
55775.11918 & $-$82.46 & $\pm$1.15 & UVES    & 3.4 \\
55775.15412 & $-$84.28 & $\pm$0.08 & UVES    & 4.8 \\
55783.09244 & $-$82.10 & $\pm$2.84 & UVES    & 3.2 \\
55783.13296 & $-$81.44 & $\pm$2.28 & UVES    & 4.6 \\
55791.22318 & $-$82.64 & $\pm$1.03 & UVES    & 6.0 \\
55791.26339 & $-$83.06 & $\pm$1.06 & UVES    & 5.7 \\
55808.06681 & $-$82.65 & $\pm$1.50 & UVES    & 6.4 \\
55808.10472 & $-$84.15 & $\pm$0.34 & UVES    & 5.9 \\
56116.40453 & $-$78.08 & $\pm$7.73 & FORS2   & 17.4  \\
56116.41831 & $-$77.25 & $\pm$6.11 & FORS2   & 15.9\\
58773.12292 & $-$82.77 & $\pm$2.27 & UVES    & 3.1 \\
\hline
\end{tabular}
\end{table}

\subsection{Spectral emission features}

\begin{figure}
\centering
\includegraphics[width=\columnwidth]{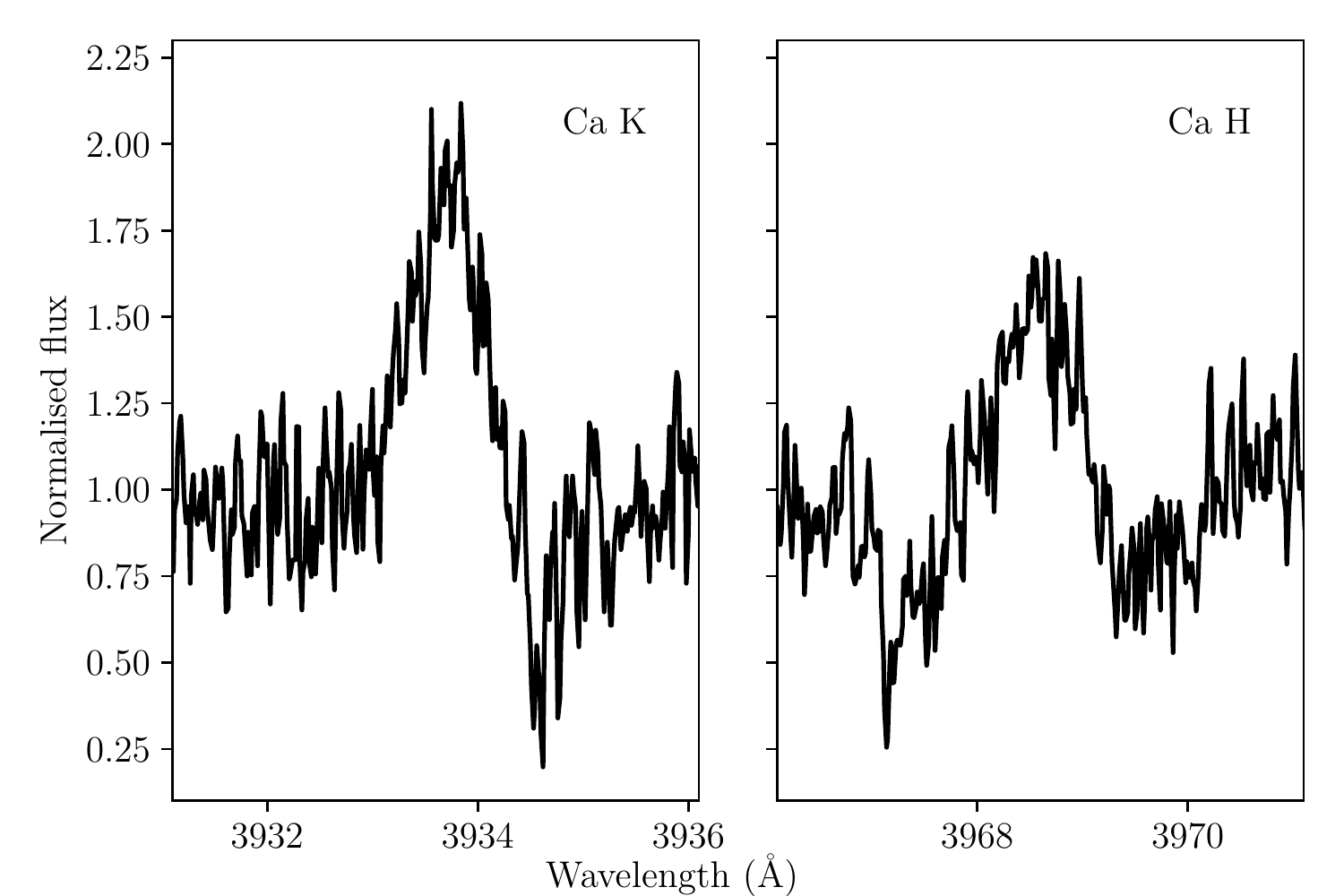}
\caption{Calcium H and K in emission in the co-added UVES spectrum of the central star of A~70.
}
\label{fig:Ca}
\end{figure}

\begin{figure}
\centering
\includegraphics[width=\columnwidth]{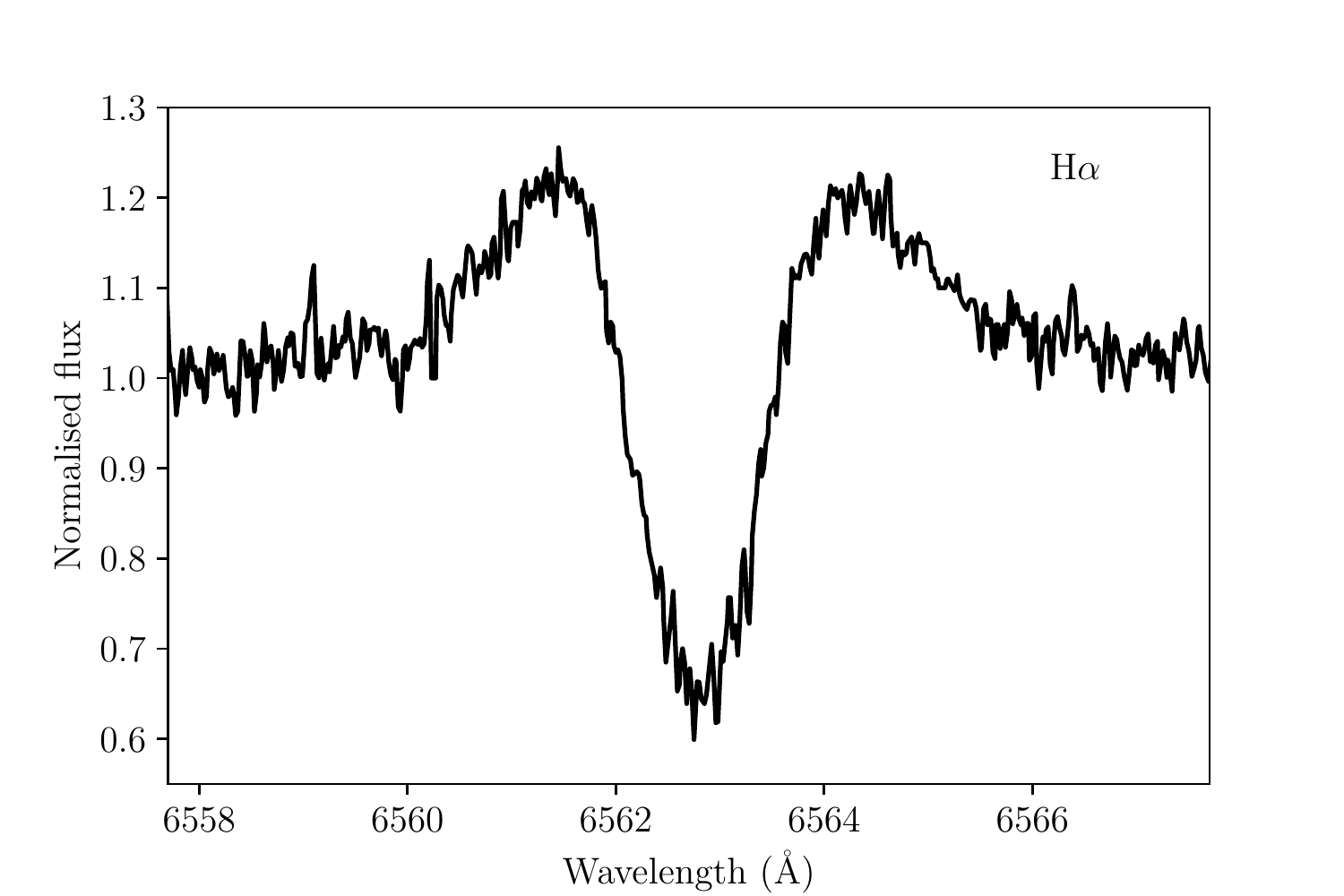}
\caption{The H$\alpha$ emission/absorption feature in the co-added UVES spectrum of the central star of A~70.
}
\label{fig:Ha}
\end{figure}

Ca H and K are in emission in the UVES spectra (see Fig.\ \ref{fig:Ca}), consistent with chromospheric activity. Similarly, just as in other rapidly rotating companions \citep[e.g.\ LoTr~1 and LoTr~5;][]{tyndall13,jasniewicz94}, H$\alpha$ shows broad emission with strong absorption superimposed  (see Fig.\ \ref{fig:Ha}).  The full-width at half-maximum of the emission ($\sim130$~\kms{}, although this is challenging to measure accurately given the strength of the absorption feature) means that it is unlikely to originate from the chromosphere of the companion but may, instead, be associated with a disc-like structure similar to those found in symbiotic stars \citep[and references therein]{tyndall13}.  Ultimately, the signal-to-noise of the spectra is insufficient to draw stronger conclusions other than to say that the central star of A~70 is spectrally very similar to others of the class.

\subsection{Spectral modelling}
\label{sec:a70model}

To determine the properties of the bright component of A70, we co-added all our spectra (after correcting for the derived radial velocities) and normalised the resulting spectrum. The \texttt{iSpec} software package \citep{ispec,ispec2}\footnote{\url{https://www.blancocuaresma.com/s/iSpec}} was then used to model the spectrum and to probe stellar parameters and abundances.  The stellar parameters ($T_\mathrm{eff}$, log $g$, $v$ sin $i$, {[M/H]}) were derived following the workflow described in \cite{ispec} using a $\chi^2$-minimization comparison with synthetic spectra calculated using the \texttt{spectrum} software \citep{spectrum} and \texttt{MARCS} model atmospheres \citep{marcs}.  The initial effective temperature and surface gravity chosen to begin the fitting process were those derived by \citet[][$T_\mathrm{eff}=5300$ K and $\log g$=3.9, respectively]{miszalski12a} with solar metallicity, and a $v$ sin $i$ consistent with the spectral type of \citet{miszalski12a} and the rotation period measured in Sec.\ \ref{sec:a70phot} (40 \kms{}, assuming an inclination, $i$, comparable to that of the nebula as determined by \citealt{tyndall13}). The fitting procedure was then allowed to undergo as many iterations as required to converge (in this case seven iterations).  

It is important to note that only iron lines (Fe \textsc{i} and Fe \textsc{ii}) were used to derive the stellar parameters, thus ensuring that any nebular contamination (or poor nebular subtraction by the UVES pipeline) did not affect the results, and that any over-abundances in s-process elements did not affect  the derivation of the metallicity (i.e., {[M/H]} $\equiv$ {[Fe/H]}).  The surface temperature ($T_\mathrm{eff}\sim$ 5000~K), surface gravity (log $g\sim3.5$) and metallicity ([Fe/H]$ \sim-0.4$) are in line with those expected based on the similarity with the G8 IV-V star HD~24616 \citep[as shown by][]{miszalski12a}\footnote{As HD~24616 is characterised by an effective temperature $T_{\rm eff}\approx 4900$~K, $\log g\approx 3.3$, and metallicity [Fe/H]$\approx -0.2$ \citep{rave}.}.  The Gaia Apsis surface gravity ($\log g\approx4.8$) is significantly larger than the determined surface gravity, but results in an appreciably poorer fit to the spectrum especially at the extremely low Apsis metallicity, [Fe/H]=$-2.73$. Here, it is also worth noting that our decision to fix the surface gravity does not dramatically impact the other parameters derived (beyond the mass) nor our subsequently derived chemical abundances. More generally, it is important to recognise that given the relatively large rotational velocity of the star, resulting in broad lines and a lot of blending, and the relatively low signal-to-noise ratio of the co-added spectrum (S/N$\sim$20), it is not possible to establish any parameters with high precision. In particular, the gravity is very poorly constrained.

\begin{figure*}
\centering
\includegraphics[width=\textwidth]{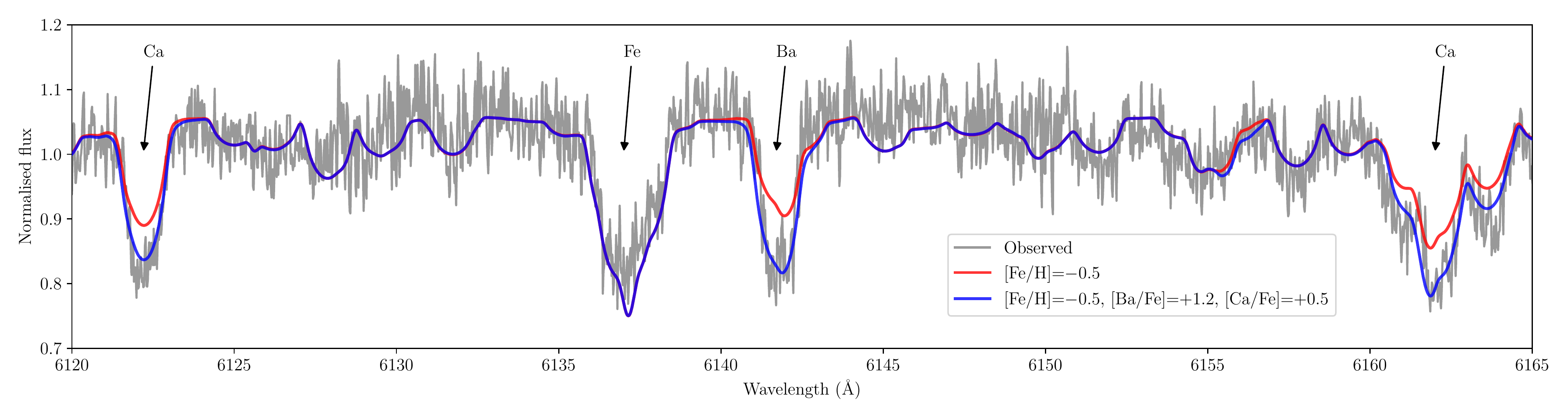}
\caption{Portion of the co-added UVES spectrum of the central star of A~70 showing the region around the 6141\AA{} Ba line (in grey). The blue line shows the best fit model with [Fe/H]=$-0.5$, [Ba/Fe]=$+1.2$ and [Ca/Fe]=$+0.5$, while the red line shows the same model without enhanced barium or calcium.
}
\label{fig:a70ba}
\end{figure*}

Using the best model, we also tried to estimate the abundance of several chemical elements using the equivalent width method, including s-process elements. Unfortunately, the abundance of only 2 elements could be derived with reasonable precision (see Fig.~\ref{fig:a70ba}): Ca and Ba, while it appears that there is no need to enhance the carbon abundance to get a good fit (nor the chromium, nickel or titanium, although these elemental abundances are less well constrained). The barium abundance is based on the barium lines at 6141 and 6496 \AA. Using the line at 4554~\AA\ results in a lower abundance of $\approx$0.8. However, it is well known that the strong line at  4554~\AA{} is severely affected by NLTE effects, which are not taken into account in our modeling, and this value should thus be considered a lower limit.  With +1 dex of Ba overabundance, the central star of A~70 is classed as a strong barium star -- these can have periods of up to 10$^4$ days but are generally found to cluster in the range of 1000--4000 days \citep{jorissen19}, compatible with the range of periods previously inferred from the UVES RVs.

The final parameters and abundances as well as their uncertainties are listed in Tab.\ \ref{tab:specparams}.

\begin{table}
\caption{Most probable range of the parameters of the bright secondary of A~70.}
\label{tab:specparams}
\centering
\begin{tabular}{ll}
\hline
Parameters & A~70\\
\hline
$T_\mathrm{eff}$ (K)			&5000$\pm$250\\
log $g$					&3.5 (fixed)\\
$v$ sin $i$ (\kms{})			& 37 $\pm$ 5\\
{[Fe/H]}					&$-0.5\pm 0.2$\\
Microturbulence  (\kms{})           & 1.0 $\pm$ 0.5 \\
{[Ba/Fe]}				 	&1.2 $\pm$ 0.2\\
{[Ca/Fe]}					& 0.5 $\pm$ 0.1\\
{[C/H]}					& $\approx$0\\
{[Ti/Fe]} & $\approx$0\\
{[Cr/Fe]} & $\approx$0\\
{[Ni/Fe]} & $\approx$0\\
\hline
\end{tabular}
\end{table}

\section{Distance, mass and radius of the companion}
\label{sec:param}

Assuming the companion rotates in the same plane as the binary orbit and the binary orbit is coincident with the waist of the nebula \citep[i.e.\ $i$=30\degr{};][]{tyndall13}, the measured rotation period of 2.06 d and $v$ sin $i$ of $\approx$37 \kms{} imply a radius of $\sim$3~R$_\odot$. For a surface gravity of log $g$=3.5, this radius would imply a mass for the companion of $\approx$1~M$_\odot$.  These parameters, as well as the previously derived temperature, are consistent with a sub-giant on the verge of joining the first giant branch \citep{pols98}.  Irrespective of the precise inclination of its rotation axis, the radius of the companion star must be $\gtrsim$1.5 R$_\odot$ (for the derived rotation period and stellar parameters).  This is still a strong indication that the star has evolved off the main sequence unless it has been inflated following the mass transfer \citep[as seen in some post-common-envelope central stars;][]{jones15,jones22} which resulted in its contamination with the products of AGB nucleosynthesis from the nebular progenitor (Ca and Ba).

Using the best-fitting synthetic spectrum from section \ref{sec:a70spec} and assuming an extinction of E(B$-$V)=0.05 \citep{frew16}, we calculated synthetic magnitudes in the three Gaia bands (see Tab.\ \ref{tab:gaia}) allowing the distance to be a free parameter.  The best-fitting distance was 4.6~kpc, in reasonable agreement with both the Gaia geometric parallax distance (3.6$^{+1.5}_{-1.0}$~kpc) and the nebular surface-brightness-radius relationship distance (6.0$\pm$2.1~kpc) of \cite{frew16}.  The Gaia Apsis distance is much closer at $\approx$1.5~kpc, however several other parameters derived by the Gaia Apsis also appear to be strongly incongruent.  While the Apsis extinction agrees well with that measured from the nebula by \citet{frew16}, the Apsis radius is not compatible with the Apsis distance and the observed Gaia magnitudes (with the radius required to match the observed photometry being approximately double the Apsis radius).  Similarly, the Apsis metallicity of [Fe/H]$\approx-$2.7 is entirely incompatible with the spectra presented in section \ref{sec:a70spec}.  The strong disagreement is hard to explain, however GSP-Phot is known to provide unreliable (and underestimated) distances for sources further than 2~kpc \citep{gaia-apsis2}, as well as to systematically underestimate metallicities \citep{andrae22}.  Ultimately, the discrepancies could be related to the GSP-Phot module's fixing of the available parameter space using isochrones, which may not be entirely compatible with a star that has experienced appreciable accretion as is the case here \citep{andrae22}.

The very strong barium enhancement we derive implies that the secondary has accreted a large amount of polluted material from the former AGB. Indeed, according to \cite{2021MNRAS.505.5554S}, one would require a dilution factor of the order of 0.25 to account for such an s-process excess. If the secondary is about 1 M$_\odot$, this means it must have accreted about 0.25 M$_\odot$ from its AGB companion. This in turn means that, assuming the primary was originally 2-3 M$_\odot$ \citep[typical for AGB progenitors of the barium star's companion;][]{karunkuzhi18}, the mass transfer efficiency must have been of the order of 10-20\% -- compatible with wind Roche lobe overflow \citep{abate13}.

\section{Primary mass}

As the primary is not visible in our spectra, we cannot place any further constraints on its temperature or radius beyond those already presented by \citet{miszalski12a}.  However, the radial velocities presented in section \ref{sec:a70spec} can be used to probe its mass.  The mass function of an eccentric binary is defined as:
\begin{equation}
    f(\mathrm{M}_1) = \frac{\mathrm{M}_1^3 \mathrm{sin}^3 i}{(\mathrm{M}_1+\mathrm{M}_2)^2} = \frac{\mathrm{P}~ \mathrm{K}_2^3}{2 \pi G} (1 - e^2)^\frac{3}{2}
\end{equation}
where M$_1$ and M$_2$ are the masses of the central star and companion, respectively. P is the orbital period, K$_2$ is the radial velocity semi-amplitude of the companion, and $e$ is the orbital eccentricity.

Assuming that the systemic velocity of the binary is, indeed, the same as the systemic velocity of the surrounding nebula, we can place limits on the majority of these parameters. The orbital period must be in the range 2 yr $\lesssim$ P $\lesssim$ 8 yr (shorter periods can be ruled out by the lack of variability in the 2011 UVES RVs), the semi-amplitude of the companion must be K$_2\gtrsim$10 \kms{} (the difference between the nebular systemic velocity and the UVES RVs), and the orbital eccentricity must be e$\gtrsim$0.3 (lower eccentricities would also not reproduce the observed lack of variability in the 2011 UVES RVs).  This leads to a mass function of 0.05--0.1~M$_\odot$ which, accounting for the derived mass of the companion (M$_2\approx$ 1~M$_\odot$) and assuming the binary orbital plane corresponds to the waist of the nebula, implies an problematically large mass for the primary (close to, or even exceeding, the Chandrasekhar mass). Stellar evolution and nucleosynthesis models indicate that the pollutors of Ba stars should almost all have had initial masses M$\sim$ 2--3 M$_\odot$ \citep{karunkuzhi18}.  This places a rough upper limit of 0.75 M$_\odot$ for the remnant based on the initial-final mass relation \citep[but with large uncertainties;][]{cummings18}.

While misalignment between rotation axis and orbital planes is not impossible, it would be particularly strange that the rotation axis of the G-type star is coincident with the nebular symmetry axis while the binary orbital plane is not. Given that the mass loss should be focused in the orbital plane one would expect that for a misaligned system it would be the binary plane that is aligned with the nebular symmetry axis rather than the stellar rotation axis.  However, such a misalignment between binary and nebula is not unheard of - there is some suggestion that the same is true for another barium star planetary nebula LoTr~5 \citep[although, given the uncertainties, there are still doubts as to whether this is proven]{jones17b,aller18}, perhaps indicating that the expected alignment between binary plane and nebular symmetry axis \citep[and references therein]{hillwig16} does not hold for such long period systems.  

It is possible that the G-type star observed at the projected centre of A~70 is simply a chance alignment and is not related to the nebula.  Indeed, other curious chance alignments have been discovered in other PNe, for example, SuWt~2 \citep{jones17c} and M~2-3 \citep{boffin18} both of which have bright main-sequence binaries close to their projected centres. However, a chance alignment is highly unlikely in the case of A~70 given that the star is contaminated in s-process elements, indicating that it must be in a binary with a more evolved companion from which it accreted s-process rich material.  Similarly, while the lack of observed radial velocity variability in the G-type star may support the chance alignment hypothesis, a similar conclusion was drawn for NGC~1514 before more extensive observations revealed that the orbit was simply too long and eccentric to have been detected in a single observing season \citep{jones17b}. 

Alternatively, the systemic velocity of the binary may not be the same as the systemic velocity of the PN. Assuming a moderate eccentricity of e$\approx$0.3 and that the orbital inclination is the same as the nebular inclination, the radial velocity semi-amplitude should be 4--5\kms{} in order for the primary mass to be approximately 0.6~M$_\odot$ (a relatively canonical value for such white dwarfs).  Based on the UVES RVs, this would mean that the binary systemic velocity should be approximately $-$79~\kms{} -- some 6~\kms{} (although only still less than two uncertainties) removed from the nebular systemic velocity of $-$73$\pm$4~\kms{} \citep{tyndall13}.

A difference in systemic velocity between binary and nebula could, in principle, be due to a ``kick'' which also helped impart eccentricity into the Ba star orbit \citep{izzard10}. In this case, the central star might be expected to be slightly offset from the centre of the nebula but, given the clumpy nature of the nebular ring, this is hard to assess using existing imagery \citep[see, e.g., Fig.\ 8 of ][]{tyndall13}.  For the same reason, it is also possible that the uncertainty on the systemic velocity of the nebula has been underestimated. There is precedent for such a conclusion with \citet{jones10a} showing that, unless a detailed spatio-kinematical modelling is performed, the brightness variations across a PN can skew the measurement of its systemic velocity by several \kms{}. This hypothesis is particularly plausible for A~70 given the clear brightness variations across the nebula where, for example, the slit number 1 of \citet{tyndall13} shows bright emission from the ring centred at a heliocentric velocity of approximately $-$80\kms{} but significantly fainter emission extending redwards to $-$30 \kms{}, while their slit position 3 shows the converse with bright emission centred at approximately $-$60 \kms{} and fainter emission extending bluewards up to $-$120 \kms{}.

Ultimately, without further observations which definitively constrain the binary period and radial velocity amplitude, it is not possible to constrain the mass of the primary, nor the possibility of misalignment between the nebular symmetry axis and binary orbital plane.

\section{Conclusions}
\label{sec:conclusions}

The rotation period of the G-type companion to the central star of A~70 is confirmed to be roughly 2~d based on photometric variability, while the spectroscopically derived parameters ($T_\mathrm{eff}$, log $g$ and $v$ sin $i$) are consistent with a G8IV-type star rotating with the same period at an inclination equal to that found for the nebula by \citet{tyndall13}.  The G-type star is chromospherically active, presenting Ca H and K in emission, consistent with having been spun-up by the accretion event which led to its enhancement in barium. A broad H$\alpha$ emission feature is visible in the spectrum of the G-type star, the origin of which is unclear but which has been observed in similar stars \citep{tyndall13}.  The spots due to the chromospheric activity are found to change phase (i.e., location on the stellar surface) in 2011, associated with a period of negligible activity, followed by a period of at least four years of stable variability.

The radial velocity measurements of the G-type star, however, do not appear to be entirely consistent with the previously derived properties of the nebula -- either indicating a higher inclination or a slightly different systemic velocity.  Both discrepancies could potentially be resolved by continued radial velocity monitoring in order to conclusively constrain the orbital period and semi-amplitude.  For now, we conclude that there is insufficient evidence to support the hypothesis that the nebula and binary are misaligned \citep[as has been suggested in another long-period central star, LoTr~5;][]{aller18}, and we encourage further detailed study of both the central star and nebula of A~70.  In particular, if the highly speculative $\sim$8 year period fit shown in Fig.\ \ref{fig:orbit} is representative, the next radial velocity maximum will occur in late-2029 -- as such this will be a critical time to obtain further radial velocity measurements.

\section*{Acknowledgments}

The authors would like to thank Howard E.\ Bond and Robin Ciardullo for making the data in their Research Note publicly available, and the anonymous referee for their insightful report.

DJ acknowledges support from the Erasmus+ programme of the European Union under grant number 2020-1-CZ01-KA203-078200. DJ  also acknowledges support under grant P/308614 financed by funds transferred from the Spanish Ministry of Science, Innovation and Universities, charged to the General State Budgets and with funds transferred from the General Budgets of the Autonomous Community of the Canary Islands by the Ministry of Economy, Industry, Trade and Knowledge. JM and AJB acknowledge the support of STFC in the form of studentships. This research was supported by the  Erasmus+ programme of the European Union under grant number 2017-1-CZ01-KA203-035562.

Based on observations made with ESO Telescopes at the La Silla Paranal Observatory under programme IDs 083.D-0654, 085.D-0629, 087.D-0174, 087.D-0205, 091.D-0475, 0103.D-0186 and 0104.D-0326.  This work makes use of observations from the LCOGT network, and observations made at the South African Astronomical Observatory (SAAO).  The CRTS survey is supported by the U.S.~National Science Foundation under grants AST-0909182 and AST-1313422. Based on observations obtained with the Samuel Oschin 48-inch Telescope at the Palomar Observatory as part of the Zwicky Transient Facility project. ZTF is supported by the National Science Foundation under Grant No. AST-1440341 and a collaboration including Caltech, IPAC, the Weizmann Institute for Science, the Oskar Klein Center at Stockholm University, the University of Maryland, the University of Washington, Deutsches Elektronen-Synchrotron and Humboldt University, Los Alamos National Laboratories, the TANGO Consortium of Taiwan, the University of Wisconsin at Milwaukee, and Lawrence Berkeley National Laboratories. Operations are conducted by COO, IPAC, and UW

This research has made use of NASA's Astrophysics Data System Bibliographic Services; the SIMBAD database, operated at CDS, Strasbourg, France; the VizieR catalogue access tool, CDS, Strasbourg, France; Astropy, a community-developed core Python package for Astronomy \citep{astropy:2013,astropy:2018}; PyAstronomy; SciPy \citep{scipy}; NumPy \citep{numpy}.

\section*{Data availability}
Individual photometric and radial velocity measurements are made available either in the paper or as online material via CDS. The raw ESO, LCOGT and LT data are available from the respective online archives, while the raw SAAO are available upon reasonable request to the corresponding author.  Survey data are available as part of their respective online data releases.

\bibliographystyle{mnras}
\bibliography{literature.bib}

\newpage
\appendix

\section{Survey photometry}
\label{sec:surveyphot}

\subsection{Catalina Real-Time Transient Survey}

\begin{figure}
\centering
\includegraphics[width=\columnwidth]{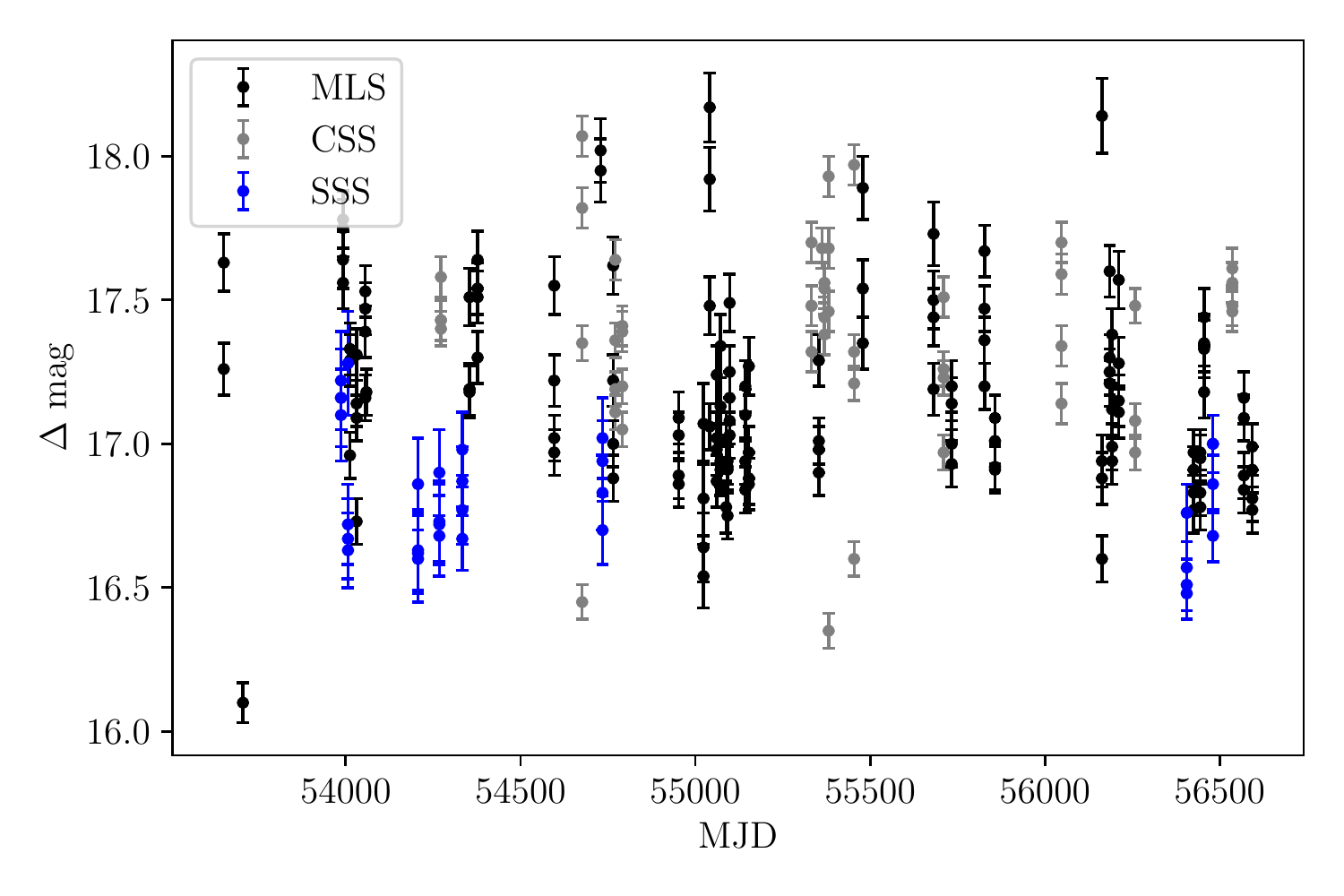}
\caption{The CRTS photometry of the central star of A~70.
}
\label{fig:CRTS}
\end{figure}

\begin{figure*}
\centering
\includegraphics[width=\columnwidth]{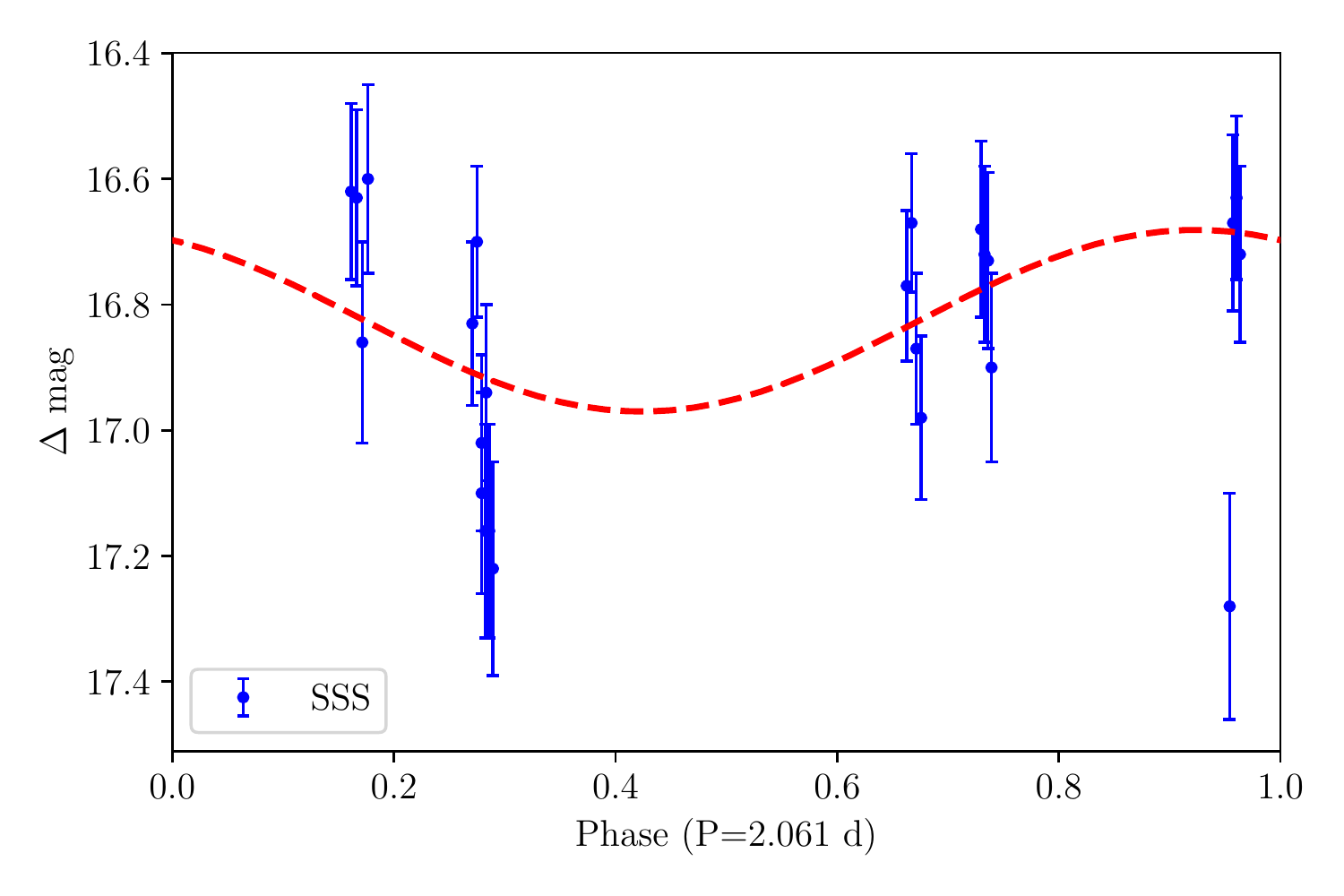}
\includegraphics[width=\columnwidth]{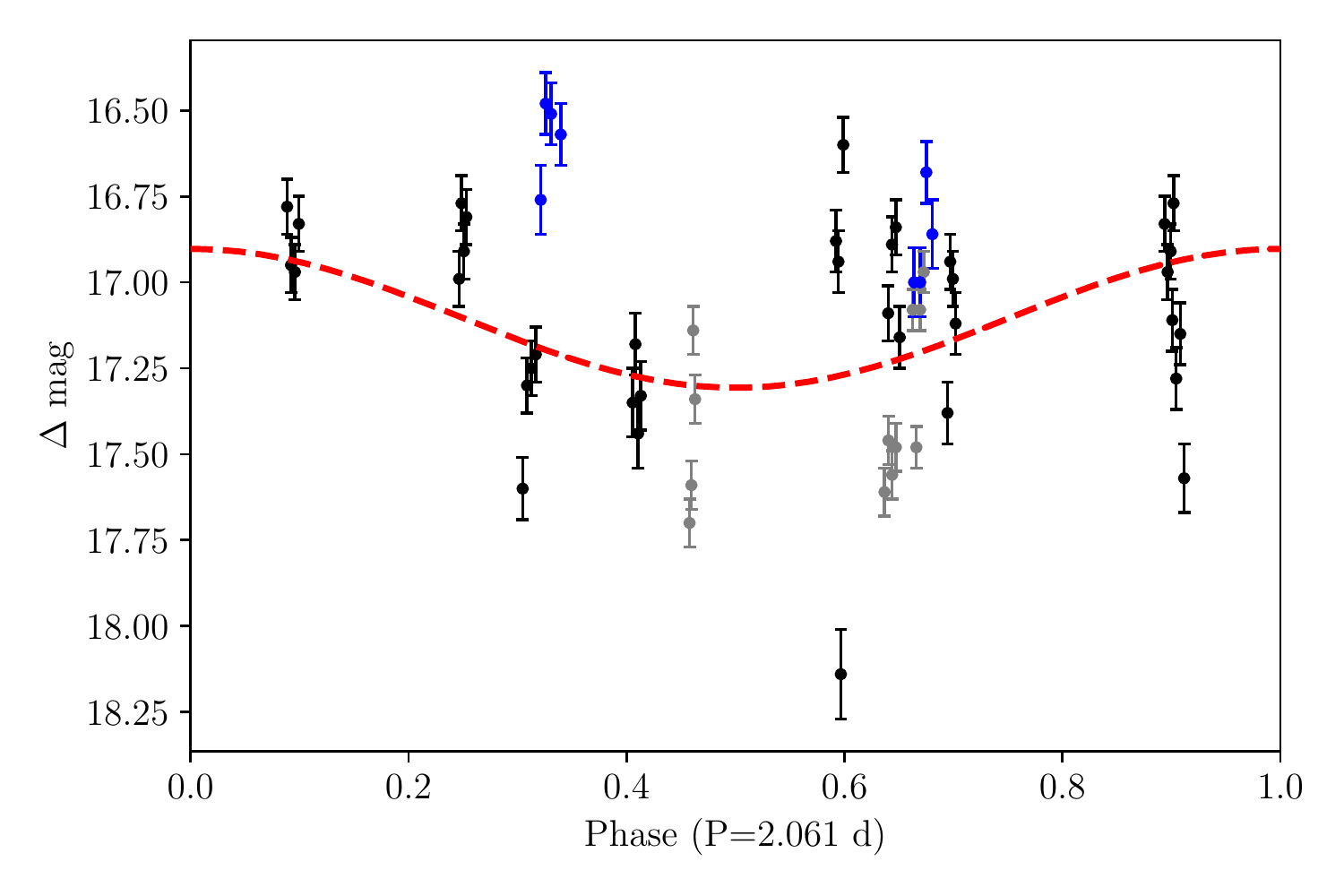}
\caption{Left: The phase-folded CRTS photometry of the central star of A~70 obtained with the SSS telescope prior to 2009. Right: The phase-folded CRTS photometry obtained after mid-2011.  All data are folded on the same ephemeris as in Fig.\ \ref{fig:a70phot}, with a sinusoidal fit shown underlaid in red.
}
\label{fig:CRTSfold}
\end{figure*}

The central star of A~70 was also observed 212 times by the Catalina Real-Time Transient Survey (CRTS) between October 2005 and November 2013 \citep{drake09}.  The combined data originating from all three telescopes used in the survey does not present with a single consistent period, with observations taken with the larger (and higher spatial resolution) 1.5m Mount Lemmon Survey (MLS) and 0.7m Catalina Sky Survey (CSS) telescopes presenting with a large ($>$1 magnitude) and aperiodic scatter.  However, the 32 points observed using the 0.5m Siding Springs Survey (SSS) telescope do present variability consistent with that derived from the photometry presented above (period $\sim$ 2 days, semi-amplitude $\sim$ 0.1 magnitudes).  Given that the photometric extraction performed by the CRTS is not optimised for the central stars of planetary nebulae, and that the nebular emission of A~70 is appreciable in the V-band employed by the survey, we conclude that the high amplitude, aperiodic variability found in the photometry from MLS and CSS are most likely spurious and related to variations in seeing \citep[and thus variable nebular contribution within the aperture;][]{jones15}.  This offers a natural explanation as to why the data from the lower spatial resolution SSS telescope is more consistent with the 2-day period derived above, as the larger pixels (pixel scale $\gtrsim$ seeing) would be less susceptible to variable nebular contamination in the aperture.  A sinusoidal fit to the early SSS data is shown in the left panel of Fig.\ \ref{fig:CRTSfold}.  In spite of the small number of measurements, the period and amplitude are seemingly consistent with the variability shown in Fig.\ \ref{fig:a70phot} but with a different phasing to the 2010 data.  This is perhaps indicative of another period of negligible variability shortly before the first observations by \citet{bond18}.  However, given the quality of the fit and small number of highly uncertain data points, we stress that this conclusion is at best speculative.  Fitting a sinusoid to all the CRTS data after mid-2011 (obtained with all three telescopes: MLS, CSS, and SSS), and assuming a 2.061~d period, reveals a best fit phasing more or less consistent with the 2011-2015 ephemeris albeit with a very large $\chi^2$ (right panel, Fig.\ \ref{fig:CRTSfold}).

\subsection{Asteroid Terrestrial-impact Last Alert System}

\begin{figure}
\centering
\includegraphics[width=\columnwidth]{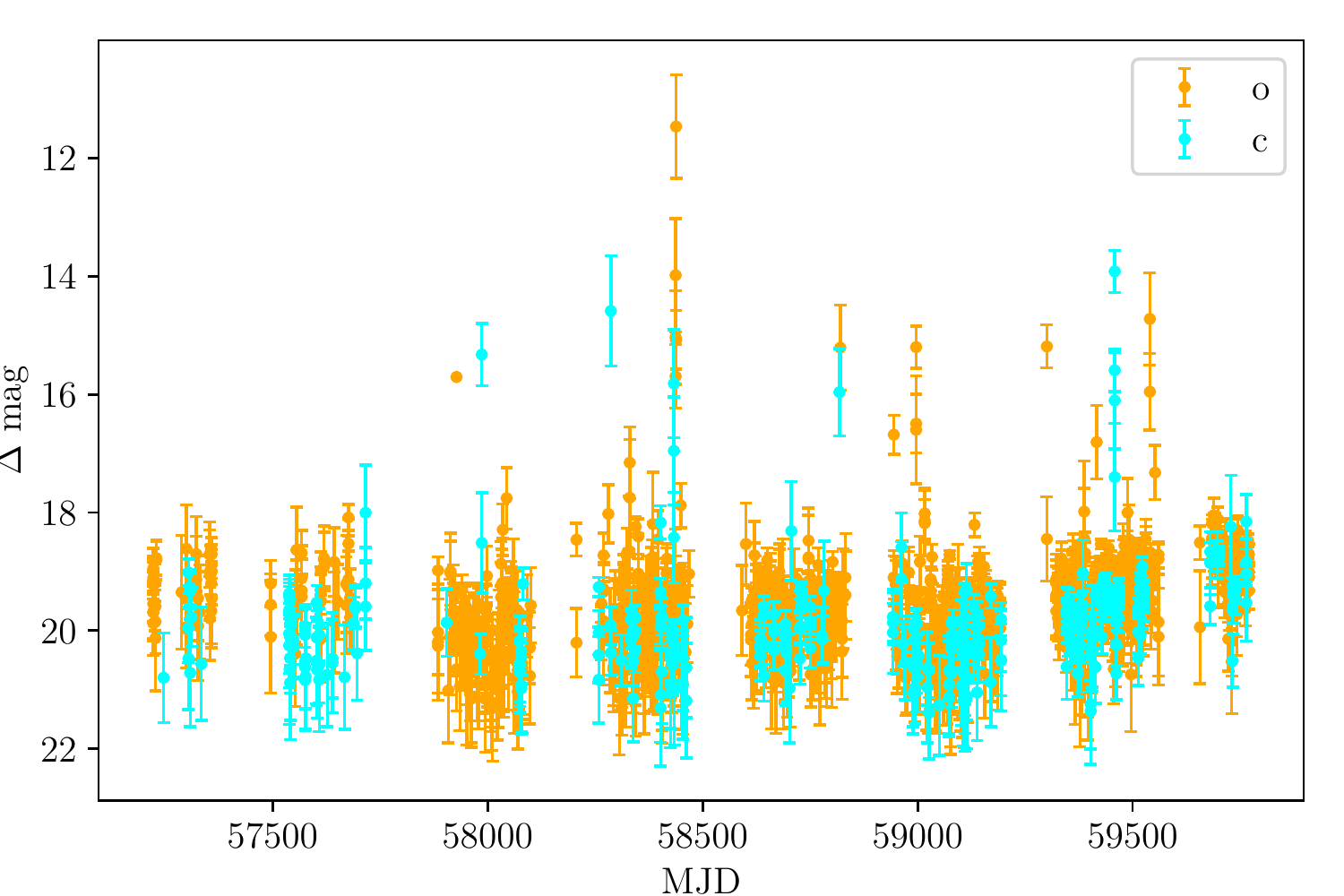}
\caption{The ATLAS photometry of the central star of A~70. Note that data points with uncertainties larger than 1 mag have been discarded.
}
\label{fig:ATLAS}
\end{figure}

The central star of A~70 was observed 1643 times in the $o$-band and 478 in the $c$-band by the Asteroid Terrestrial-impact Last Alert System \citep[ATLAS;][]{ATLAS} between July 2015 and July 2022. The data are rather imprecise for our purposes, with a median uncertainty in both bands of approximately 0.6 magnitudes (see Fig.\ \ref{fig:ATLAS}). Perhaps unsurprisingly, there are no significant peaks in the periodograms and the data do not phase well on the previously measured photometric period.

\subsection{All-Sky Automated Survey for Supernovae}

\begin{figure}
\centering
\includegraphics[width=\columnwidth]{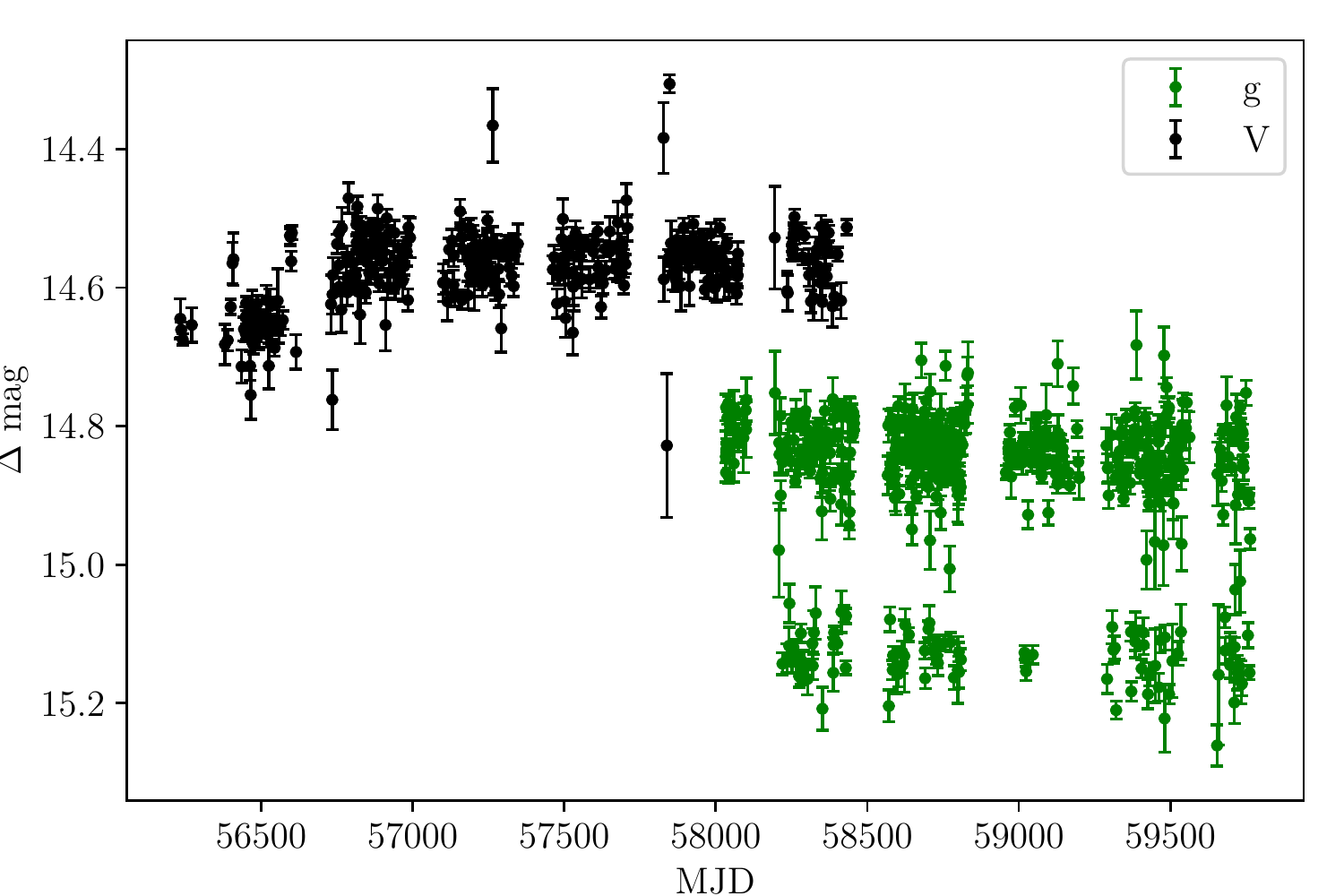}
\caption{The ASAS-SN photometry of the central star of A~70.
}
\label{fig:ASASSN}
\end{figure}

The central star of A~70 was observed by the All-Sky Automated Survey for Supernovae \citep[ASAS-SN;][]{asas-sn,asas-sn2} 338 times in the $V$-band between November 2012 and November 2018, and a further 517 times in the $g$-band between October 2017 and July 2022.  The magnitudes from ASAS-SN are significantly brighter ($\sim$4 magnitudes) than the values from both CRTS and ATLAS (see Fig.\ \ref{fig:ASASSN}), strongly indicative that the photometry is heavily contaminated by the surrounding nebula.  Furthermore, the $g$-band data separates in to two separate regimes, similarly indicative of an issue with the photometry.  Sinusoid fitting to the $V$-band data results in a phasing consistent with the 2011-2015 ephemeris, but with an amplitude that is too low (only 0.01 mag) and a very large reduced $\chi^2$ ($>$10).  As one might expect from their bi-modal nature, the $g$-band data do not phase well with any ephemeris.  Therefore, just as for the ATLAS data, the ASAS-SN data are not useful in constraining the variability of the central star of A~70.

\subsection{Zwicky Transient Facility}

\begin{figure}
\centering
\includegraphics[width=\columnwidth]{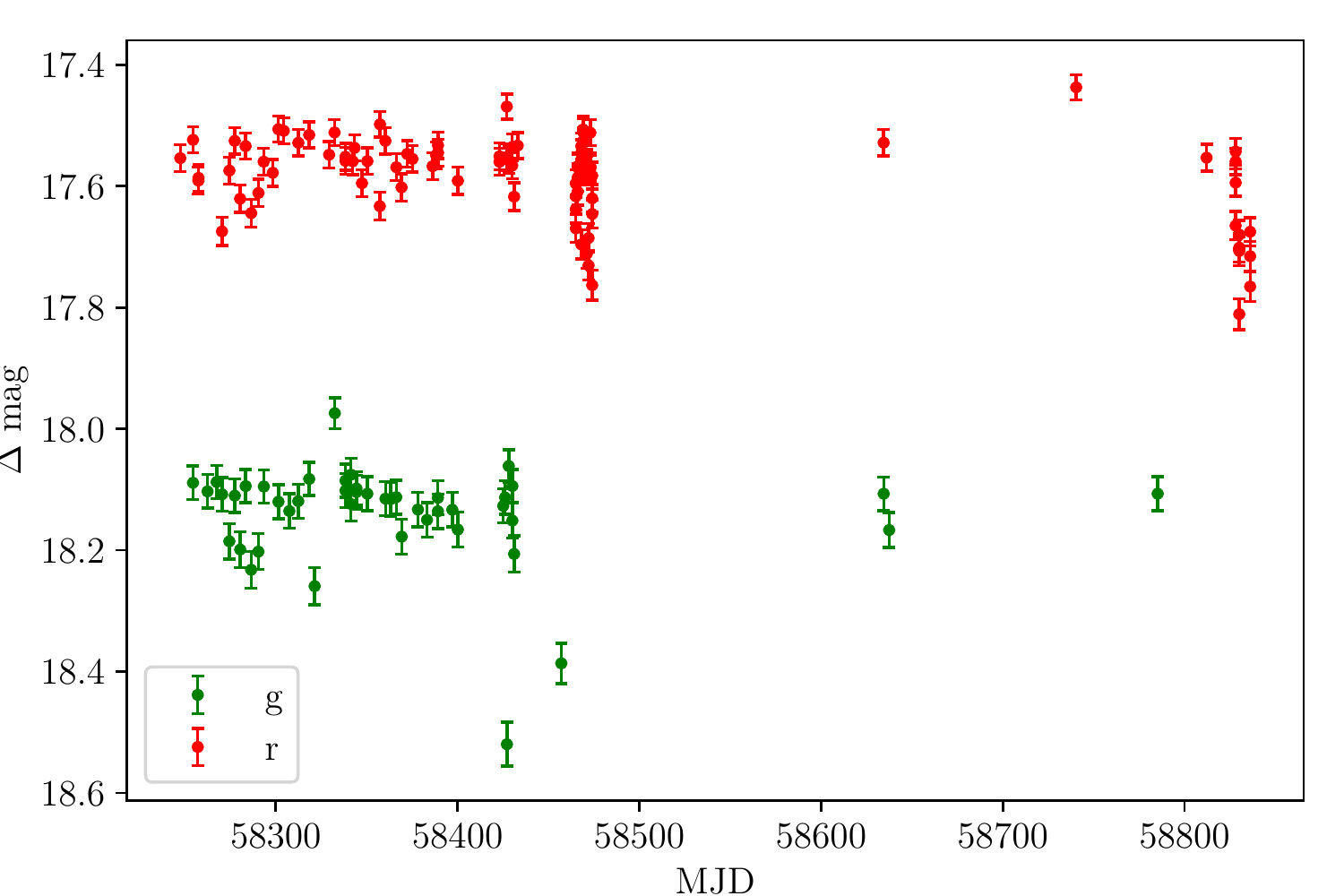}
\caption{ZTF photometry of the central star of A~70.
}
\label{fig:ZTF}
\end{figure}

\begin{figure}
\centering
\includegraphics[width=\columnwidth]{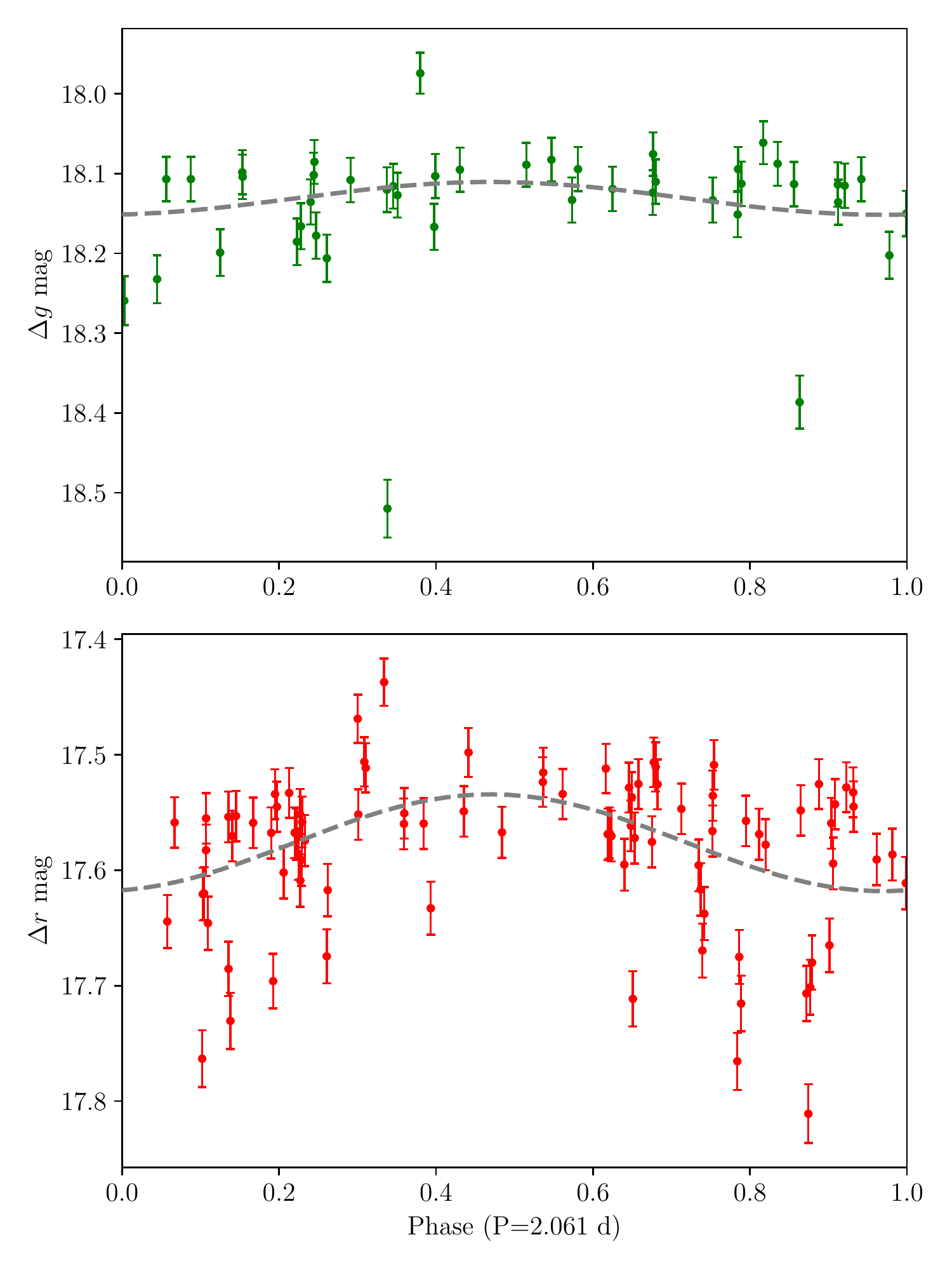}
\caption{Phase-folded ZTF photometry of the central star of A~70. The sinusoidal fit to the data (assuming the same time of minimum but different amplitudes) are shown underlaid in grey. 
}
\label{fig:ZTFfold}
\end{figure}

The central star of A~70 was observed 45 times in the $g$-band and 90 times in the $r$-band by the Zwicky Transient Facility \citep[ZTF;][]{ztf} between May 2018 and December 2019.  Very much like the CRTS data, the ZTF data shows a relatively large ($\sim$0.4 magnitudes) and seemingly aperiodic scatter in both bands (see Fig.\ \ref{fig:ZTF}).  Both show peaks in their periodograms at periods close to 2.06~d (although neither is significant).  Simultaneous sinusoid fitting of both $g$- and $r$-bands, as shown in Fig.\ \ref{fig:ZTFfold}, results in a phasing which is not consistent with the 2011-2015 and/or 2022 data presented in Fig.\ \ref{fig:a70phot} (but rather more consistent with the 2010 epoch).  While not conclusive (given the uncertainty on the photometric period), this is seemingly indicative that the variability was not stable over the intervening time period.

\subsection{Transiting Exoplanet Survey Satellite}

The field containing A~70 was observed by the Transiting Exoplanet Survey Satellite \citep[TESS;][]{TESS} in Sector 54 (9 July 2022 -- 5 August 2022), however at the time of submission that data has yet to be released.  In any case, at an approximate TESS magnitude of 17.1 \citep[as listed in the TESS input catalog;][]{TIC}, it is highly unlikely that TESS will return useful photometry of the central star system.  Furthermore, the TESS pixel size of 21\arcsec{} is comparable to the angular size of the nebula, so the TESS photometry of the central star is likely be heavily contaminated by nebular emission.

\label{lastpage}

\end{document}